\documentclass{article}

\usepackage{graphicx}
\usepackage{color} 
\usepackage{amssymb} 
\usepackage{lineno}
\usepackage{natbib}

\makeatletter
\renewcommand{\@seccntformat}[1]{}
\makeatother

\bibpunct{}{}{,}{s}{}{}

\begin{document}

\title{ The mysterious cut-off of the 
Planetary Nebula Luminosity Function
}
\author{K. Gesicki$^*$, A. A. Zijlstra, and M. M. Miller Bertolami}
\date{}
\maketitle

\section{Letter to {\em Nature Astronomy}}


{\bf Planetary Nebulae (PNe) mark the end of the active life of 90\%\
  of all stars. They trace the transition from a red giant to a degenerate white
  dwarf.  Stellar models\cite{1994ApJS...92..125V, 1995A&A...299..755B}
  predicted that only stars above approximately twice the solar mass
  could form a bright PN.  But the ubiquitous presence of bright PNe
  in old stellar populations, such as elliptical galaxies, contradicts
  this: such high mass stars are not present in old systems.  The planetary
  nebula luminosity function (PNLF), and especially its bright
  cut-off, is almost invariant between young spiral galaxies, with
  high mass stars, and old elliptical galaxies, with only low mass
  stars. Here we show that new evolutionary
  tracks of low-mass stars are capable to explain 
  in a simple manner
  the decades-old mystery. 
  The agreement between the observed PNLF and 
  stellar evolution validates the latest theoretical modelling. With
  these models, the PNLF provides a powerful diagnostic to derive star
  formation histories of intermediate-age stars.
  The new models predict that the Sun at the end of its life
  will also form a PN, but it will be faint.}


Low and intermediate mass stars, up to about 8\,$M_\odot$, end their
lives with a phase of extreme mass loss. The superwind ejects the
envelope, leaving only the degenerate core behind. The core briefly
ionizes the ejecta, before entering the terminal white dwarf cooling
stage. The ionized ejecta remain visible as a planetary nebula (PN)
for thousands of years, before dispersing into the interstellar 
medium\cite{2013A&A...558A..78J}.

PNe form the most luminous phase of evolution of their host stars,
with typical luminosity $L\sim 10^4 L_\odot$. In addition, much of the
nebular luminosity comes out in a few bright, narrow emission
lines. The brightest line, [O\,III]\,5007\,\AA, can emit as much as
$10^3\,L_\odot$. This makes PNe detectable to very large distances,
including galaxies beyond the Virgo cluster.  The PNLF has been
established as an important extragalactic distance
estimator\cite{1988ASPC....4...42J, 2012Ap&SS.341..151C}. However,
stellar evolution models have thus far been unable to explain
it\cite{2016IAUFM..29B..15C}.

The PNLF describes the fraction of PNe in a galaxy at each specified
luminosity bin\cite{1980ApJS...42....1J}. For the emission line
[O\,III]\,5007\,\AA, the absolute magnitude is
defined\cite{1973asqu.book.....A} as
$M_{5007} = -2.5 \log(F[{\rm O\,III}]) -13.74$. Here
$F([{\rm O\,III}])$ is the line flux in units of
erg\,s$^{-1}$\,cm$^{-2}$, assuming a distance of 10\,pc. In these
units, the PNLF shows a well-defined, steep cut-off at the bright end,
at $M^*_{5007}\simeq-4.5$.  The observed value of $M^*_{5007}$ is
almost invariant between different galaxies, with a scatter of only
$\sim 0.17$\,mag\cite{2012Ap&SS.341..151C} for metal-rich galaxies.
The observed magnitudes include the effect of extinction within the
circumstellar material of the PN, which especially for very young and
compact PNe can be of the order $\sim 0.5$\,mag.

To reproduce the invariant cut-off 
{ some studies\cite{1997A&A...321..898M} assumed } 
a final stellar mass
distribution with an upper cut-off at 0.63\,$M_\odot$ (initial mass:
$M_{\rm i}\approx 2\,M_\odot$). This is much too high for older
stellar populations\cite{2008ApJ...681..325M}.  
{ Other studies\cite{2004A&A...423..995M, 2007A&A...473..467S} }
showed that the stellar evolution models available before
2016\cite{1994ApJS...92..125V, 1995A&A...299..755B}, combined with
nebular models of evolving optical thickness, predicted a brightening
of $M^*_{5007}$ by more than 4 magnitudes when going from old
(10\,Gyr) to young (1\,Gyr) stellar populations, in stark
contradiction with the observed invariance of $M^*_{5007}$.
These inconsistencies are a
long standing mystery in the study of the PNLF, and have thrown doubts
on our understanding of the final evolution of low mass stars,
including the Sun.

\begin{figure}[t]
\centering
\includegraphics[width=10cm]{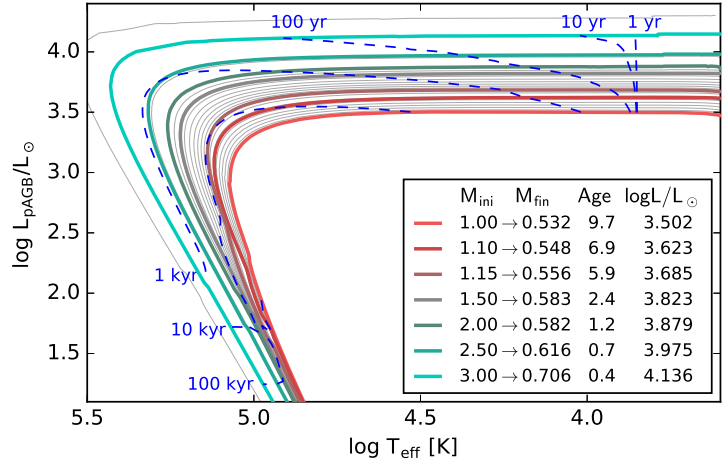}
\caption{ {\bf Stellar evolution sequences and timescales.}  The coloured
  lines present the post-AGB stellar evolution
  sequences\cite{2016A&A...588A..25M} adopted for the computations of
  the fluxes shown in Fig.\,\ref{o3evolu}. Blue dashed lines indicate
  isochrones at selected post-AGB ages. Thin grey lines show the
  interpolated sequences used for the construction of the PNLFs shown
  in Fig.\,\ref{histall}. The inset presents the model parameters for
  the seven computed tracks: the initial and final masses in $M_\odot$
  units, the total age since ZAMS in Gyr and the $\log L/L_\odot$.}
 \label{stellarevolu}
\end{figure}

Recently, the details of the stellar evolution models
\cite{1994ApJS...92..125V, 1995A&A...299..755B} applied in the
modelling of the PNLF have been questioned. The analysis of PNe in the
Galactic bulge\cite{2014A&A...566A..48G} suggested that the speed of
the stellar evolution during the PN phase had been underestimated by a factor of
three or more.  Independently, new stellar evolution models for
post-asymptotic giant branch (post-AGB) stars were
developed\cite{2016A&A...588A..25M}, with a carefully computed
post-AGB phase based on new opacities, an updated treatment of the
stellar physics and a proper account of previous evolutionary
stages (see Fig.\,\ref{stellarevolu}). These new models have 
very different initial-final mass relations 
{ calibrated to be in a good agreement with observations.}
Their post-AGB evolution is faster and brighter than the
previous models. The new models fundamentally change the
interpretation of the PNLF.

From this new model grid, we selected seven evolutionary sequences
with metallicity $Z=0.01$, representative for a wide range of
Solar-like populations. The initial masses range from 1 to
3\,$M_\odot$ and the final masses from 0.532 to 0.706\,$M_\odot$
respectively. This initial mass range corresponds to stellar ages
between 0.5 and 10\,Gyr (see the inset in
Fig.\,\ref{stellarevolu}). The grid includes previously published
sequences\cite{2016A&A...588A..25M} and newly computed models using
the same physical assumptions.

The evolutionary sequence provides the temperature evolution and the
luminosity evolution of the post-AGB star.  To model the PNLF we
calculate emission line fluxes of model nebulae surrounding each one
of the stellar models.  The complete photoionization structures of the
nebulae were calculated using the Torun
codes\cite{1996A&A...309..907G}, to derive the flux of the single
nebular line used, [O\,III]\,5007\,\AA.

We first ran models with a non-evolving constant density shell, with a
fixed inner radius of 0.01\,pc, outer radius of 0.02\,pc, and a mass
of 0.15\,$M_\odot$.  These parameters were chosen to result in
ionization bounded (opaque) nebulae and were adopted for each of the
seven evolutionary sequences.  In this maximum-nebula hypothesis, the
fluxes emitted at each phase of the stellar evolution are
maximized. It is appropriate for testing whether these PNe can
reproduce the peak of the PNLF.

\begin{figure}[t]
\centering
\includegraphics[width=10cm]{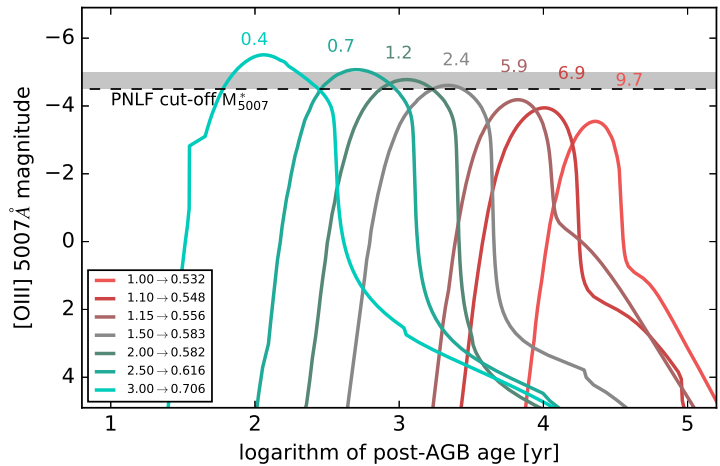}
\caption{{\bf Evolution of [O\,III]\,5007\,\AA\ fluxes
  against logarithmic time for the maximum-nebula hypothesis}. 
  The solid lines represent
  evolutionary tracks of central stars, with initial and final masses
  as given in the inset. The numbers above each curve indicate the
  total age in Gyr of a given stellar model. The PNLF canonical
  cut-off value of  $M_{5007} = -4.5$\,mag is shown as a horizontal dashed line;
  the grey band represents the possible range of 
  this value when correction for extinction was considered.}
 \label{o3evolu}
\end{figure}

For each evolutionary sequence we computed the fluxes emitted in the
line [O\,III]\,5007\,\AA, converted to the magnitude scale $M_{5007}$.
Fig.\,\ref{o3evolu} shows this magnitude evolution during the post-AGB
evolution for the seven low-mass stars introduced in
Fig.\,\ref{stellarevolu}.  The most massive track reaches magnitudes
above the observed PNLF cut-off value. However the time spent at
maximum brightness is negligible in comparison with the average
life-time of PNe: the typical observed kinematic age of PNe is between
$\lesssim 1000$ and 10,000\,yrs.  In addition, the high-mass, youngest
PNe are expected to be more compact and may have significant
circumstellar extinction.
The next five less-massive tracks behave very regularly and
over thousands of years they show an [O\,III] brightness very close to
the observed cut-off value, exceeding it (if at all) by only a
fraction of a magnitude. This is a consequence of the very similar
post-AGB luminosity of these models (see the inset in
Fig.\,\ref{stellarevolu}) which changes by only 0.25 dex between the
1.1\,$M_\odot$ and the 2\,$M_\odot$ sequences (ages 7\,Gyr and 1\,Gyr
respectively).  Finally, the least massive track, at 1\,$M_\odot$
evolves so slowly that the ejected nebula will be dispersed before the
star culminates in $T_{\rm eff}$.  The new stellar evolution models
coupled with photoionization modelling under the maximum-nebula
hypothesis predict the maximum nebular [O\,III] brightness to be close
to the observed value over a large range of stellar ages and masses
(Fig.\,\ref{o3evolu}). This already suggests that the models can
reproduce the PNLF cut-off for a variety of stellar
populations.

Next, we allow the modelled gaseous shell to evolve together with
the central star. We assumed again a constant density shell with
fixed inner radius of 0.01\,pc and fixed total mass of
0.10\,$M_\odot$, but now the outer radius expands from 0.02\,pc up
to 0.5\,pc at a constant velocity. The expansion of the shell 
results in a transition of the
nebula from opaque to transparent to the ionizing radiation. Varying
the kinematics provides different [O\,III] brightness scenarios. In
the following we will discuss two such scenarios: one where the PN
is predominantly opaque during the stellar evolution (the
intermediate-nebula hypothesis) and the other one where the PN is
predominantly transparent, the minimum-nebula hypothesis. 

\begin{figure}
\centering
\includegraphics[width=9cm]{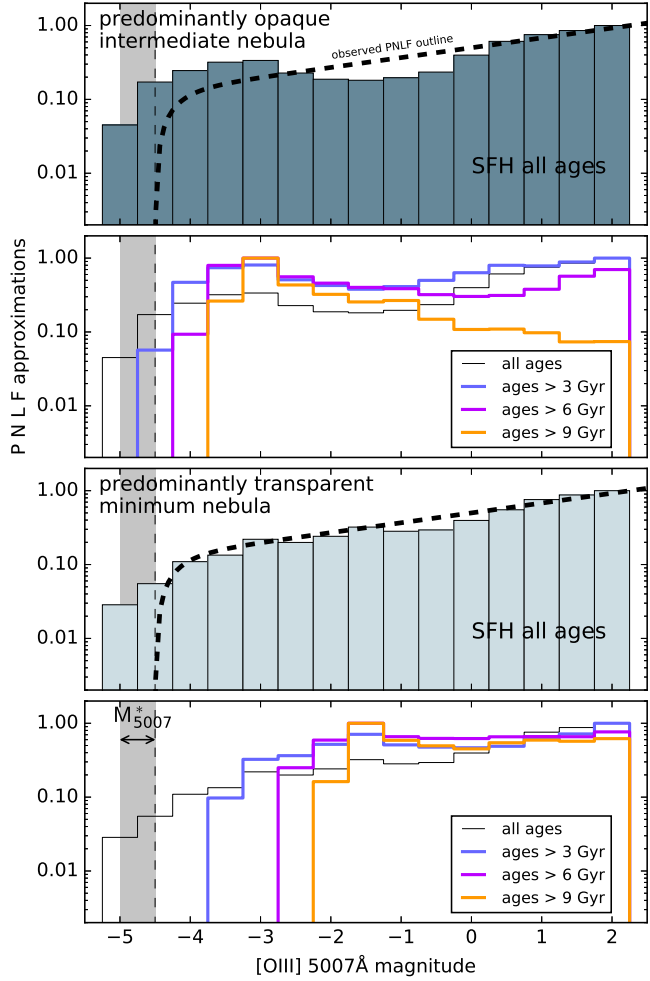}
\caption{{\bf The synthesized PNLF for the intermediate-nebula 
    and minimum-nebula hypothesis, for different star
    formation histories.}  We interpolated between evolutionary tracks
    (Fig.\,\ref{stellarevolu})
    to obtain a grid of models sampled at uniform age. For each assumed
    SFH we combined the individual PNLFs from each gridded model, with
    weighting factors determined by the SFH, the mass difference between
    adjacent interpolated tracks, and assuming a Salpeter initial mass
    function. The PNLF synthesis was performed for two versions of
    a very simple nebular evolution scenarios. The upper two panels
    presents the PNe that stay opaque for most of the post-AGB
    evolution (expansion of the shell starts later and with lower
    velocity), the bottom two panels concern PNe that become transparent 
    relatively early (expansion starts earlier and with higher
    velocity). The shaded histograms show the PNLF obtained with a
    continuous SFH since some 10\,Gyr ago until now.  The coloured
    lines show simulations with star formation ending 3, 6, and 9 Gyr
    ago. Both shaded histograms are supplemented with the observed
    PNLF outline that combines\cite{1989ApJ...339...53C} the bright
    steep end and the faint tail. At the left side of all panels the
    canonical value of $M^*_{5007}$ is indicated as a vertical dashed
    line with a grey band representing the possible range of 
    this value when correction for extinction was considered. }
 \label{histall}
\end{figure}

To derive the PNLF, we integrate over the stellar mass distribution,
assuming a Salpeter initial mass function (IMF) with
$n(M) \propto dM \times M^{-2.35}$.  We interpolate between the
stellar models (Fig.\,\ref{stellarevolu}) to get a grid of tracks 
at equally spaced ages.
Fig.\,\ref{histall} combines the models to predict PNLFs for both
the intermediate and minimum-nebulae hypothesis, and for four
different star formation histories (SFHs), appropriate to different
galaxy types. The filled histograms (dark for the 
intermediate-nebula hypothesis with mostly opaque PNe, pale for the 
minimum-nebula hypothesis with more transparent PNe) use
a constant SFH,
approximating a non-interacting spiral galaxy. The other two
diagrams (shown with coloured lines) use a truncated SFH where 
a constant star formation ceased 3, 6 and 9 Gyr ago, typical for an elliptical
galaxy and for an old stellar population, formed in a burst about 9
Gyr ago.

For the intermediate-nebula hypothesis (the two upper panels in
Fig.\,\ref{histall}), the steep bright-end cut-off of the PNLF appears in our
simulations for both the continuous and truncated star formation
histories, at a value very close to the canonical observed limit of
$M^*_{5007}\sim -4.5$.  The brightest PNe are created by progenitors
1.1\,$M_\odot<M_{\rm i}<2\,M_\odot$, which have nearly the same
post-AGB stellar luminosity,
$\log ( L_{\rm pAGB}/L_\odot ) = 3.75\pm 0.13$, and have total ages
from 1 to $\sim$7\,Gyr.  The initial stellar masses are significantly
smaller than was required in { similar }
older models. { The fact that the cut-off 
appears at the right magnitude 
over a wide age range of stellar populations provides a simple solution to} 
the long standing mystery of the observed PNLF cut-off invariance.

The two lower panels in Fig.\,\ref{histall} present the
minimum-nebula hypothesis. It reproduces the observed faint-tail
behaviour of PNLF, however the bright-end becomes progressively fainter 
with age of the population. 

We conclude that the PNLF 
can be reproduced using the new stellar models, under the assumptions
that the ejected shells can be approximated with the
intermediate-nebula hypothesis, and that the originating stellar population
{ harbours a significant number of stars with}
ages of 3--7\,Gyr. The age range is consistent with the known SFH,
where nearby galaxies with masses between $10^{10}$ and
$10^{11}\,M_\odot$ tend to have declining star formation rates since
3--5\,Gyr ago\cite{2004Natur.428..625H}.  { Elliptical
galaxies also can contain a fraction of stars in this age
range\cite{2015MNRAS.448.3484M}.  Our results confirm previous
studies\cite{1997A&AS..122..215R} which concluded that appropriate
PNLF peak luminosities could be attained by lower-mass stars if the
nebulae remained optically thick during most of their evolution. }

{ 
The necessary condition that the modelled spherical PNe remain
opaque during a significant part of their early evolution 
can perhaps be satisfied by a high opacity in one direction
and transparency in another. }
For bipolar PNe containing a dense equatorial torus and
thin lobes, 3D modelling\cite{2016A&A...585A..69G} reproduces well the
line emissivities (including [O\,III]). {  The details
of the evolution and expansion of PNe are still not well
understood. Our results puts constraints on this evolution, for
those PNe which make up the peak of the PNLF.  }

The brightness and the evolutionary speed of the central stars of PNe,
both crucial to the PNLF, are influenced by physical processes on the
previous AGB phase, in particular the mass loss and the mixing
which bring the products of nuclear burning to the
surface. The details of the observed PNLF may put constraints on the
efficiency of these still poorly understood processes in AGB stars.

The models give new information about the final fate of the Sun.
In our models, it is near  the low-mass limit for PN formation. The Sun
still reaches a temperature high enough to ionize the ejecta, before
they disperse. Hydrogen ionization begins
5000 years after the end of the AGB, and [O\,III] appears from 12,500
years.  The stellar luminosity suffices to put its PN within a
magnitude of the bright cut-off of the PNLF.  However, at this late
time, it is expected that the ejecta will no longer be optically
thick, and therefore the PN may be rather fainter than
this. If the Sun leaves the AGB during the
helium burning phase of the thermal pulse cycle, its luminosity and
speed of evolution will be three times lower and no PN will
form. The Sun is close to the lowest mass star that can still produce a PN.




\vspace{0.5cm}

\section{Methods}


\subsection{Extinction corrections to the PNLF cut-off}

The extinction towards extragalactic PNe is a combination of foreground extinction
(within our Galaxy), local extinction (within  the host
galaxy) and circumstellar extinction (within the stellar ejecta).  The
foreground is normally constant across the host galaxy and is known. The local
extinction is a  {  stochastic} parameter, which 
mostly affects PNe located within the disk of spiral galaxies. 
Circumstellar extinction stems from dust located in
the non-ionized regions surrounding the PN.  {
Observed PNLF are generally corrected for foreground extinction, but
not for the poorly known local and circumstellar extinction.}

We calculate the circumstellar extinction using the equation \cite{2003Olofsson}
$$A_{\rm V} \approx 0.5 \biggl[ {\dot M \over 10^{-5}\,\rm
      M_\odot\,yr^{-1} }\biggr]  
     \biggl[ {v_{\rm exp} \over 15 \rm \, km\, s^{-1} } \biggr]^{-1}
     \biggl[  { r\over 10^{16}\,\rm cm} \biggr]^{-1} .
$$
This assumes a standard ratio of
$N_{\rm H_2} = 2 \times 10^{21} A_{\rm V} \,\rm cm^{-2}\,mag^{-1}$, appropriate for
interstellar silicate dust \cite{2009MNRAS.400.2050G}.  For our minimum radius
of  0.01\,pc and
assuming $\dot M = 5 \times 10^{-5}\,M_\odot\, {\rm yr}^{-1}$ and $v_{\rm
  exp} = 15\,\rm km\,s^{-1}$, this predicts $A_{\rm V} \approx 0.8\,\rm
mag$. The smallest radii of Galactic PNe are around 0.03\,pc \cite{2016MNRAS.455.1459F}, giving an expected $A_{\rm V}\approx 0.25\,$mag.  

\begin{figure}[t]
\centering
\includegraphics[width=10cm]{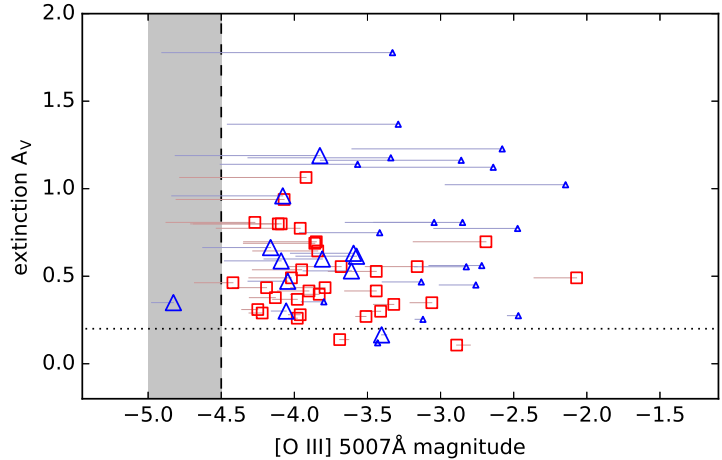}
\caption{ {\bf Observed values of $M_{5007}$ for
planetary nebulae of the { outer disk} 
of the Andromeda galaxy, M\,31.}   
The data is taken from different sources, shown as red 
squares\cite{2012ApJ...753...12K, 2015ApJ...807..181C, 2015ApJ...815...69F, 2013ApJ...774....3B}
and blue triangles\cite{2014AJ....147...16K}.
Extinctions were calculated from the published
H$\alpha$/H$\beta$ ratio, assuming a ratio of 2.85. 
$M_{5007}$ was taken from a single source\cite{2006MNRAS.369..120M}, 
with one exception\cite{2014AJ....147...16K} 
where $M_{5007}$ is measured from the spectra itself. The small
triangles indicate objects  
where the extinction is less accurate because of low S/N on the H$\beta$ line. 
The foreground extinction is $A_{\rm V}=0.175$, 
extinction at [O\,III] is taken as $1.2\,A_{\rm V}$, 
and the distance modulus to M\,31 is taken as 24.47. 
{ The coloured symbols show $M_{5007}$ corrected for foreground
extinction; the thin lines indicate the $M_{5007}$ magnitudes after correction for the total extinction. }
The canonical value of $M^*_{5007}=-4.5$ is shown as the vertical dashed line. 
{ The grey vertical band represents the corresponding range of
values assuming extinction correction up to 0.5 mag.
The horizontal dotted line represents the foreground
component to $A_{5007}$.}
}
 \label{extinc}
\end{figure}

As a test, the well studied PNe population of the { outer disk} 
of the Andromeda galaxy M\,31 is shown in Fig.\,\ref{extinc}. 
The foreground extinction to M\,31
amounts to $A_{\rm V}=0.175$. 
The corresponding value of $A_{5007}$ is presented as the horizontal dotted line. 
{ The coloured symbols show  $M_{5007}$  corrected
for foreground extinction only, and the thin lines indicate the shift
when corrected for the total extinction, which here is derived from
the spectra. }
The majority of objects have additional 
extinctions of a few tenths of a magnitude or less. A few have higher
extinction. 
{ After correcting for the full extinction, 
the value of the cut-off of the PNLF shows a small shift from
$M^*_{5007}=-4.2$ to $-4.5$. In our models, the former value would
indicate an age of the stellar population of around 7\,Gyr, while the
corrected value corresponds to stellar ages of less than 6\,Gyr. This
illustrates the diagnostic power of the PNLF to constrain
star-formation histories, provided accurate extinctions are available.
In this age range,  few other diagnostics
are available. It should however be noted that extinction
corrections can be uncertain, especially if there
is significant circumstellar extinction and scattering, potentially 
involving non-standard dust grains\cite{2012MNRAS.422..955G}.

The grey band in the plots shows the dimming effect a circumstellar extinction of up to 0.5 mag would have on the observed PNLF, when compared to the theoretical, extinction-free curves. Such an extinction is possible for the most compact PNe, from the above calculations.
}

\subsection{La Plata stellar evolution code}

LPCODE is a one-dimensional stellar evolution code that has been
widely used for the computation of full evolutionary sequences
from the zero age main sequence to the white dwarf stage
\cite{2006A&A...454..845M, 2010ApJ...717..183R,
2013A&A...557A..19A}.  The last version  of
LPCODE\cite{2016A&A...588A..25M}  includes a state-of-the-art
treatment of atomic, molecular and conductive opacities
\cite{1996ApJ...464..943I, 2007ApJ...661.1094C,
2009A&A...508.1343W}  as well as a detailed treatment of stellar
winds and convective boundary mixing. Specifically, LPCODE takes
into account both for carbon- and oxygen-rich compositions for
the determination of both opacity and the stellar wind regime.
In addition, convective boundary mixing has been carefully
calibrated at different evolutionary stages to reproduce several
AGB and post-AGB observables\cite{2016A&A...588A..25M}. 
The post-AGB evolution of the seven tracks computed with this code 
and discussed in this article are presented in
Fig.\,\ref{stellarevolu}. 
Each star evolves first at constant luminosity to
higher temperatures, followed by a rapid fading and slow cooling. The
speed of evolution is much faster at higher stellar mass, so that
lower-mass stars spend much longer at their peak luminosity than
higher mass stars do. 
{A metallicity of $Z=0.01$ was selected, as being
appropriate for sun-like stars.  The post-AGB tracks have little
dependence on metallicity, and using $Z=0.02$ would cause only minor
changes to the derived masses and ages of the stars.
A 1.25\,$M_\odot$ track\cite{2016A&A...588A..25M} was not used 
for the interpolations because it was affected
by a thermal pulse unusually close to the AGB turn-off which 
affected the subsequent evolution.
}
The 1\,$M_\odot$ presented here
was adjusted to ensure it left the AGB  during the hydrogen burning
phase, by briefly turning off the stellar wind until the hydrogen
shell had fully activated.

\subsection{Torun photoionization code}

The Torun codes were written in their original form some twenty
years ago\cite{1996A&A...309..907G}. The central star is
approximated as a black-body. The nebula is approximated with a
spherically symmetric shell having assumed inner and outer radii and radial density
distribution. The chemical composition of the nebular gas
was taken from the data set named PLANetary which is included as a part
of the CLOUDY code\cite{2013RMxAA..49..137F}. No dust presence
was assumed. The ionization state of gas was computed by solving
the equations of statistical equilibrium and iterated to
convergence with the kinetic temperature and electron density.
These codes were not written in a form suitable for public
availability but they are easy to adapt to different needs and
run very fast on standard workstations. They have been used  in
numerous publications.

\subsection{[O\,III] magnitudes for individual stellar tracks}

We first derived [O\,III] fluxes from a single evolutionary track, for a
particular stellar mass. 
At this step we were interested in maximizing the flux emitted in
the [O\,III] line. The shell parameters were adopted to secure that
the PN stays opaque during the whole stellar evolution, the
maximum-nebula hypothesis. As the stellar temperature increases and
the number of ionizing photons goes up, the ionization front expands
and the ionized mass increases. 
To obtain opaque PNe, the total mass of the shell should
exceed the ionized mass. We found that a constant density shell of
inner radius of 0.01\,pc, outer radius of 0.02\,pc and  total mass
of 0.15\,$M_\odot$ fulfils our demands for each of the central star tracks. 
The resulting hydrogen number density $\sim 1.5 \cdot 10^5$cm$^{-3}$ 
guarantees confinement of the ionization front even for the hottest stars.
These parameters remain typical for unevolved PNe. 
For each of the tracks presented in Fig.\,\ref{stellarevolu}
the flux evolution in the [O\,III] line was
calculated using the Torun model described above.
The [O\,III] line flux we converted to magnitude scale and
in this way the dependencies shown in Fig.\,\ref{o3evolu} were obtained. 

We derived also the central star luminosity fraction that is
reprocessed into nebular flux in [O\,III]\,5007\,\AA\ line.
The maxima of this reprocessing efficiency remain nearly constant, 
at the value $\sim 0.12$ which is in agreement with earlier modelling\cite{2010A&A...523A..86S}.
However the actual values of this efficiency vary significantly with 
stellar evolution, firstly they increase with $T_{\rm eff}$ 
then start to decrease because of higher oxygen ionization 
and further decrease because the star enters the cooling track. 
In fact they follow closely the time dependencies of [O\,III]\,5007\,\AA\ fluxes 
drawn in Fig.\,\ref{o3evolu}.

\subsection{PNLF for expanding PNe in composite stellar populations}

To account for simultaneous evolution of the central star and the
  expansion of nebular shell we applied a simple scheme. We kept the
  nebular inner radius fixed and the total nebular mass constant. The
  shell is assumed to have a constant density with radius. We
  specified the outer radius as expanding with a constant
  velocity. The expansion of the outer radius together with a constant
  inner radius agrees with the observed velocity fields, which have
  the highest expansion velocities at the outer edge and very low
  velocity near the inner radius. The expansion results in a fast
  decrease of the gas density and in consequence of the nebular
  opacity. The model with expansion starting at the moment when the
  star leaves the AGB almost immediately becomes transparent.
  Hydrodynamic calculations\cite{2013A&A...558A..78J} show a brief
  phase of acceleration, caused by the overpressure of the developing
  ionized region, followed by a plateau. We therefore introduced a
  delay of the beginning of the (constant) expansion phase. As
  previously we want to apply the same shell model to all of the
  synthetic tracks. Because the evolutionary tracks differ
  significantly in timescales we apply the temperature criterion --
  the nebula starts to expand after the central star effective
  temperature reaches a specified value. It does not mean that the
  early phase is static -- as the star heats up the ionization front
  moves through the opaque shell increasing gradually the nebular
  ionized radius and the ionized mass. Two parameters define our
  simplistic models: the threshold temperature which defines the early
  opaque phase and the expansion velocity which determines the late
  transparent phase of nebular evolution. As previously the inner
  radius equals 0.01\,pc while the outer radius initially has the
  value of 0.02\,pc and later expands but not more than to
  0.5\,pc. The total mass to be ionized was assumed 0.1\,$M_\odot$,
it is a little lower than at previous computations since now we are interested
in nebulae that at some moment should become transparent. 
The resulting hydrogen number densities decrease with post-AGB time
from  $\sim 10^5$ to $\sim 5$cm$^{-3}$. 
  All these parameters agree with the rather wide limits
  of physical characteristics that define a PN\cite{2010PASA...27..129F}.

  With trial and error we pinpointed two nebular models that represent
  two representative scenarios. One model is predominantly opaque for
  all evolutionary tracks: it starts expanding with velocity of
  20\,km\,s$^{-1}$ after the star heats-up to 60,000\,K. The second
  model expands a little faster (30\,km\,s$^{-1}$) and expansion
  starts somewhat earlier at the temperature of 40,000\,K, this model
  is predominantly transparent for all simulated tracks. These
  expansion velocities are typical to PNe. The two models are
  respectively the intermediate-nebula and the minimum-nebula hypothesis.

  The most massive stellar model ($M_{\rm i}=3\,M_\odot$) in both
  hypotheses evolves so fast that the nebula stays opaque throughout
  the complete brightest phase and becomes transparent only very late
  on the cooling track. Also for the least-massive track
  ($M_{\rm i}=1\,M_\odot$) the nebula stays opaque but for the opposite
  reason -- the temperature evolution is very slow. For the five
  intermediate evolutionary tracks the "opaque period" varies from 600
  to 6000\,years for the minimum-nebulae hypothesis,  while for the
  intermediate-nebula hypothesis it remains opaque about 50\%\
  longer. In both cases, the periods for
  the nebulae for staying opaque are within reasonable limits.

The carefully computed set of seven tracks (see Fig.\,\ref{stellarevolu})
served as the base for interpolation on a dense time-grid. 
In order to interpolate among the tracks we
identified key points corresponding to well-defined post-AGB
evolutionary phases.  Then, the individual evolutionary tracks have
been reduced to the same number of ``equivalent points'' between these
key points. For a given age we then compute the expected mass of the
post-AGB remnant ($M^{\rm synthetic}_{\rm f}$) and construct the
interpolated track by interpolating in final mass at equivalent points
between the two neighbouring tracks -- with masses $M^{\rm evol.,
  A}_{\rm f}$ and $M^{\rm evol., B}_{\rm f}$ such that $M^{\rm evol.,
  A}_{\rm f}<M^{\rm synthetic}_{\rm f}<M^{\rm evol., B}_{\rm f}$.
We computed
synthetic tracks interpolated to a uniform grid in the age of the
stellar population, to every 0.5\,Gyr in the range from 0.25\,Gyr to
9.75\,Gyr. 
This grid forms the base for assembling simple stellar populations.

To build a histogram we divided the magnitude axis into 0.5\,mag
bins (a value comparable with published observed data). The relative
time spent by the evolutionary track within given magnitude limits
provides a measure of the probability of finding a PN in this bin.
We interpolated the evolutionary track to a dense time grid and
counted the number of such grid-points within each magnitude bin.
A step of 5\,yr appears sufficient to obtain an adequate PNLF. The
post-AGB evolution slows down significantly with decreasing stellar
luminosity leading to more numerous faint objects. 
After some time the PN disperses. We calculate the time 
spent in each magnitude bin up to the adopted by us 20,000 years limit
which is well above the ages that contribute to our histograms
and is in agreement with recent PN visibility analysis\cite{2013A&A...558A..78J}.

Having this data set we can integrate the PNLF contributions
from individual synthetic tracks where the SFH and IMF are used to
calculate weighting factors for each track. We simulate a simple,
constant SFH started and finished at the given ages. 
However the equal steps in stellar age translate to steps $dM$
in initial mass $M$ that are increasing with $M$. To account for the
IMF we therefore multiplied each individual contribution by the mass
step $dM$ and by the Salpeter exponent $M^{-2.35}$ and then summed
them up. In this way the histograms in Fig.\,\ref{histall} were
obtained separately for opaque and for transparent shells.

\newpage


\newpage

\section{Acknowledgements}

AAZ and KG acknowledge the financial support by The University of
Manchester and by Nicolaus Copernicus University.  AAZ is supported
by the UK Science and Technology Facility Council (STFC) under grant ST/P000649/1.
M3B is partially suported by ANPCyT and CONICET through
grants PICT-2 014-2708 and PIP 112-200801-00940 and also by a Return
Fellowship from the Alexander von Humboldt Foundation.

\section{Author informations}

\subsection{Affiliations}

\noindent K. Gesicki\\
Centre for Astronomy, Faculty of Physics, Astronomy and Informatics, 
Nicolaus Copernicus University, Grudziadzka 5, PL-87-100 Torun, Poland; \\
e-mail: kmgesicki@umk.pl\\ \\
A. A. Zijlstra\\
Jodrell Bank Centre for Astrophysics, School of Physics \&\ Astronomy, \\
University of Manchester, Oxford Road, Manchester M13\,9PL, UK; \\
Department of Physics \&\ Laboratory for Space Research, University of
Hong Kong, Pok Fu Lam Road, Hong Kong;\\
e-mail: a.zijlstra@manchester.ac.uk\\ \\
M. M. Miller Bertolami\\
Instituto de Astrof\'isica de La Plata, UNLP-CONICET, \\
Paseo del Bosque s/n, 1900 La Plata, Argentina; \\
e-mail: mmiller@fcaglp.unlp.edu.ar

\subsection{Contributions}

AAZ and KG developed the concept. 
MMMB provided the post-AGB evolutionary sequences obtained with LPCODE 
and computed the supplementary data. 
KG adopted the Torun codes for the present work, performed the 
photoionization calculations, and synthesized the PNLF. 
All authors participated in discussions of the results, 
their presentations in figures and descriptions in manuscript 
and in pinpointing the conclusions.

\subsection{Competing interests}

The authors declare no competing financial interests.

\subsection{Corresponding author}

K.Gesicki


\begin{thebibliography}{10}

\bibitem{1994ApJS...92..125V}
{Vassiliadis},~E., and {Wood},~P.~R.
\newblock {Post-asymptotic giant branch evolution of low- to intermediate-mass
  stars}.
\newblock {\em ApJS} {\bf 92,} 125--144 (1994).

\bibitem{1995A&A...299..755B}
{Bloecker},~T.
\newblock {Stellar evolution of low- and intermediate-mass stars. II. Post-AGB
  evolution.}
\newblock {\em A\&A} {\bf 299,} 755 (1995).

\bibitem{2013A&A...558A..78J}
{{Jacob}, R., {Sch{\"o}nberner}, D. and {Steffen}, M.}
\newblock {The evolution of planetary nebulae. VIII. True expansion rates and visibility times.}
\newblock {\em A\&A} {\bf 558,} 78 (2013)

\bibitem{1988ASPC....4...42J}
{Jacoby}, G.~H., {Ciardullo}, R., and {Ford}, H.~C.
\newblock {Planetary nebulae as distance indicators}.
\newblock {\em Astronomical Society of the Pacific Conference Series}, {\bf 4,} 42--56 (1988).

\bibitem{2012Ap&SS.341..151C}
{Ciardullo},~R.
\newblock {The Planetary Nebula Luminosity Function at the dawn of Gaia}.
\newblock {\em Ap\&SS} {\bf 341,} 151--161 (2012).

\bibitem{2016IAUFM..29B..15C}
{Ciardullo},~R.
\newblock {The Planetary Nebula Luminosity Function and its Issues}.
\newblock {\em  Proceedings of the IAU} {\bf 29B,} 15-19 (2016).

\bibitem{1980ApJS...42....1J} 
Jacoby, G.~H.,
\newblock{The luminosity function for planetary nebulae and the number of
planetary nebulae in local group galaxies.}
\newblock{{\em ApJS}, {\bf 42,} 1-18 (1980)}

\bibitem{1973asqu.book.....A}
{Allen}, C.~W.,
\newblock {Astrophysical quantities}.
\newblock {{\em London: University of London, Athlone Press}, 3rd ed. (1973). Formula on page 197.}

\bibitem{1997A&A...321..898M}
{Mendez},~R.~H., and {Soffner},~T.
\newblock {Improved simulations of the planetary nebula luminosity function.}
\newblock {\em A\&A} {\bf 321,} 898--906 (1997).

\bibitem{2008ApJ...681..325M}
 {M{\'e}ndez},~R.~H.,  {Teodorescu},~A.~M., {Sch{\"o}nberner},~D., {Jacob},~R., and
  {Steffen},~M.
\newblock {Toward Better Simulations of Planetary Nebulae Luminosity
  Functions}.
\newblock {\em ApJ} {\bf 681,} 325--332 (2008).

\bibitem{2004A&A...423..995M}
{Marigo},~P., {Girardi},~L., {Weiss},~A.,  {Groenewegen},~M.~A.~T., and {Chiosi},~C.
\newblock {Evolution of planetary nebulae. II. Population effects on the bright
  cut-off of the PNLF}.
\newblock {\em A\&A} {\bf 423,} 995--1015 (2004).

\bibitem{2007A&A...473..467S}
{Sch{\"o}nberner},~D., {Jacob},~R., {Steffen},~M., and {Sandin},~C.
\newblock {The evolution of planetary nebulae. IV. On the physics of the
  luminosity function}.
\newblock {\em A\&A} {\bf 473,} 467--484 (2007).

\bibitem{2014A&A...566A..48G}
{Gesicki},~K., {Zijlstra},~A.~A., {Hajduk},~M., and {Szyszka},~C.
\newblock {Accelerated post-AGB evolution, initial-final mass relations, and
  the star-formation history of the Galactic bulge}.
\newblock {\em A\&A} {\bf 566,} A48 (2014).

\bibitem{2016A&A...588A..25M}
{Miller Bertolami},~M.~M.
\newblock {New models for the evolution of post-asymptotic giant branch stars
  and central stars of planetary nebulae}.
\newblock {\em A\&A} {\bf 588,} A25 (2016).

\bibitem{1996A&A...309..907G}
{Gesicki},~K., {Acker},~A., and {Szczerba},~R.
\newblock {Modelling the structure of selected planetary nebulae.}
\newblock {\em A\&A} {\bf 309,} 907--916 (1996).

\bibitem{2004Natur.428..625H} 
{Heavens},~A., {Panter},~B., {Jimenez},~R., and {Dunlop},~J.
\newblock{The star-formation history of the Universe from the stellar populations of nearby galaxies}
\newblock {\em Nature}, {\bf 428,} 625-627 (2004)

\bibitem{2015MNRAS.448.3484M} 
{McDermid},~R.~M., {Alatalo},~K., {Blitz},~L., {et al.}
\newblock{The ATLAS3D Project - XXX. Star formation histories and stellar population scaling relations of early-type galaxies}
\newblock{MNRAS}, {\bf 448,} 3484 (2015)

\bibitem{1997A&AS..122..215R}
{Richer}, M.~G., {McCall}, M.~L. and {Arimoto}, N.
\newblock {Theoretical models of the planetary nebula populations in galaxies: The ISM oxygen abundance when star formation stops}.
\newblock {\em A\&AS} {\bf 122,} 215--233 (1997).

\bibitem{2016A&A...585A..69G}
{Gesicki}, K., {Zijlstra}, A.~A. and {Morisset}, C.
\newblock {3D pyCloudy modelling of bipolar planetary nebulae: Evidence for fast fading of the lobes}
\newblock {\em A\&A} {\bf 585,} A69 (2016).

\bibitem{1989ApJ...339...53C}
{{Ciardullo}, R., {Jacoby}, G.~H., {Ford}, H.~C. and {Neill}, J.~D.},
\newblock{Planetary nebulae as standard candles. II - The calibration in M\,31 and its companions.}
\newblock{\em ApJ} {\bf 339,} 53--69 (1989)

\bibitem{2012ApJ...753...12K} 
Kwitter, K.~B., Lehman, E.~M.~M., Balick, B., \& Henry, R.~B.~C., 
\newblock{Abundances of Planetary Nebulae in the Outer Disk of M31}
\newblock{\em ApJ}, {\bf 753,} 12 (2012)

\bibitem{2015ApJ...807..181C} 
Corradi, R.~L.~M.,  Kwitter, K.~B., Balick, B., Henry, R.~B.~C., \& Hensley, K.,
\newblock{The Chemistry of Planetary Nebulae in the Outer Regions of M31}
\newblock{\em ApJ}, {\bf 807,} 181 (2015)

\bibitem{2015ApJ...815...69F} 
Fang, X., Garc{\'{\i}}a-Benito, R., Guerrero, M.~A., et al.,
\newblock{Chemical Abundances of Planetary Nebulae in the Substructures of M31}
\newblock{\em ApJ}, {\bf 815,} 69 (2015)

\bibitem{2013ApJ...774....3B} 
Balick, B., Kwitter, K.~B., Corradi, R.~L.~M., \& Henry, R.~B.~C.,
\newblock{Metal-rich Planetary Nebulae in the Outer Reaches of M31} 
\newblock{\em ApJ}, {\bf 774,} 3 (2013)

\bibitem{2014AJ....147...16K} 
Kniazev, A.~Y.,   Grebel, E.~K., Zucker, D.~B., et al., 
\newblock{A Search for Planetary Nebulae with the Sloan Digital Sky Survey: The Outer Regions of M31}
\newblock{\em AJ}, {\bf 147,} 16  (2014)

\bibitem[Merrett et al.(2006)]{2006MNRAS.369..120M} Merrett, H.~R.,
  Merrifield, M.~R., Douglas, N.~G., et al.,
\newblock{A deep kinematic survey of planetary nebulae in the Andromeda galaxy using the Planetary Nebula Spectrograph}
\newblock{\em MNRAS}, {\bf 369,} 120 (2006)


\bibitem{2003Olofsson} Olofsson, H,
\newblock{Circumstellar Envelopes},
in: {\em Asymptotic Giant Branch Stars} (H.J. Habing, H. Olofsson, eds.)
(Springer, Berlin, 2003) 

\bibitem{2009MNRAS.400.2050G} G{\"u}ver,
  T., \& {\"O}zel, F.,
\newblock{The relation between optical extinction and hydrogen column density in the Galaxy} 
\newblock{\em MNRAS}, {\bf 400,} 2050 (2009)

\bibitem{2016MNRAS.455.1459F} Frew, D.~J., Parker,
  Q.~A., \& Boji{\v c}i{\'c}, I.~S.,
\newblock{The Hα surface brightness-radius relation: a robust statistical distance indicator for planetary nebulae} 
\newblock{\em MNRAS}, {\bf 455,} 1459 (2016)

\bibitem{2012MNRAS.422..955G} 
{Gray},~M.~D., {Matsuura}, M., \& {Zijlstra},~A.~A.\ 
\newblock{Radiation transfer in the cavity and shell of a planetary nebula}
\newblock{MNRAS}, {\bf 422,} 955  (2012)

\bibitem{2006A&A...454..845M}
{Miller Bertolami},~M.~M., and  {Althaus},~L.~G.
\newblock {Full evolutionary models for PG 1159 stars. Implications for the
  helium-rich O(He) stars}.
\newblock {\em A\&A} {\bf 454,} 845--854 (2006).

\bibitem{2010ApJ...717..183R}
{Renedo},~I., et al.
\newblock {New Cooling Sequences for Old White Dwarfs}.
\newblock {\em ApJ} {\bf 717,} 183--195 (2010).

\bibitem{2013A&A...557A..19A}
{Althaus}~L.~G., {Miller Bertolami},~M.~M., and  {C{\'o}rsico}~A.~H.
\newblock {New evolutionary sequences for extremely low-mass white dwarfs.
  Homogeneous mass and age determinations and asteroseismic prospects}.
\newblock {\em A\&A} {\bf 557,} A19 (2013).

\bibitem{1996ApJ...464..943I}
{Iglesias},~C.~A., and  {Rogers},~F.~J.
\newblock {Updated Opal Opacities}.
\newblock {\em ApJ} {\bf 464,} 943 (1996).

\bibitem{2007ApJ...661.1094C}
{Cassisi},~S.,  {Potekhin},~A.~Y., {Pietrinferni},~A., {Catelan},~M., and
  {Salaris},~M.
\newblock {Updated Electron-Conduction Opacities: The Impact on Low-Mass
  Stellar Models}.
\newblock {\em ApJ} {\bf 661,} 1094--1104 (2007).

\bibitem{2009A&A...508.1343W}
{Weiss},~A., and  {Ferguson},~J.~W.
\newblock {New asymptotic giant branch models for a range of metallicities}.
\newblock {\em A\&A} {\bf 508,} 1343--1358 (2009).

\bibitem{2013RMxAA..49..137F}
 {Ferland},~G.~J., et al.
\newblock {The 2013 Release of Cloudy}.
\newblock {\em Rev. Mexicana Astron. Astrofis.} {\bf 49,} 137--163 (2013).

\bibitem{2010A&A...523A..86S}
{Sch{\"o}nberner}, D., {Jacob}, R., {Sandin}, C. and {Steffen}, M.,
\newblock {The evolution of planetary nebulae. VII. Modelling planetary nebulae of distant stellar systems}.
\newblock {\em A\&A} {\bf 523,} A86 (2010).

\bibitem{2010PASA...27..129F}
{{Frew}, D.~J. and {Parker}, Q.~A.},
\newblock{Planetary Nebulae: Observational Properties, Mimics and Diagnostics}
\newblock{\em PASA} {\bf 27,} 129--148 (2010)

\end{thebibliography}
\end{document}